\documentclass{olplainarticle}
\usepackage{subfigure}
\usepackage{caption}
\usepackage{hyperref}
\usepackage{xcolor}
\usepackage{soul}
\usepackage{natbib}

\title{Curvature of an Arbitrary Surface for Discrete Gravity and for $d=2$ Pure Simplicial Complexes}

\author[1]{Ali H. Chamseddine}
\author[1,2]{Ola Malaeb}
\author[1,2]{Sara Najem}
\affil[1]{Department of Physics, American University of Beirut, Beirut, Lebanon}
\affil[2]{Center for Advanced Mathematical Sciences, American University of Beirut, Beirut, Lebanon}
\keywords{Discrete gravity, Curvature, Networks, Simplicial Complexes, Computational Geometry}

\begin{abstract}
We propose a computation of curvature of arbitrary two-dimensional surfaces of three-dimensional objects, which is a contribution to discrete gravity with potential applications in network geometry.  We begin by linking each point of the surface in question to its four closest neighbors, forming quads.  We then focus on the simplices of $d=2$, or triangles embedded in these quads, which make up a pure simplicial complex with $d=2$. This allows us to numerically compute the local metric along with zweibeins, which subsequently leads to a derivation of discrete curvature defined at every triangle or face. We provide an efficient algorithm with $\mathcal{O}(N \log{N})$ complexity that first orients two-dimensional surfaces, solves the nonlinear system of equations of the spin-connections resulting from the torsion condition, and returns the value of curvature at each face.  
\end{abstract}

\begin{document}

\flushbottom
\maketitle
\thispagestyle{empty}

\section{Introduction}

In general relativity and, naturally, in differential geometry, the curvature of a continuous manifold or that of spacetime is of crucial importance, since they are tightly related to gravity. Discrete counterparts of curvatures have been proposed, such as Regge for discrete manifolds \cite{bauer2011ollivier,ollivier2007ricci,jost2014ollivier,sreejith2016forman,regge1961general}, in the context of discrete general relativity from which the continuum limit could be recovered. This definition of curvature is associated with each face of with a $d$-dimensional discrete manifold. It depends on the dihedral angles of the $d$-dimensional simplices incident to the face \cite{bianconi2021higher}.
Lately, another definition of curvature of discrete spaces was introduced in \cite{discretegravity}, where the discrete manifolds were developed using elementary cells (with Planck volume). Each cell is defined by its own tangent space and displacement operators, allowing the movement from a given cell to those which it shares a common  boundary with.  These cells are characterized by a finite number of operators and spin connections. It was shown that as the volume of the elementary cells vanishes, the standard results for the continuous curved differentiable manifolds are recovered, without relying on dihedral angles as is the case for Regge curvature. In \cite{discrete1}, \cite{discrete2}, and \cite{blackhole}, computational methods were used to find the curvature for different surfaces in two- and three-dimensions. 

In the current paper, we propose a method that generalizes to an arbitrary two-dimensional surface and provides a numerical implementation of the procedure. The advantage of such a discrete gravity theory \cite{discretegravity} is that a finite number of degrees of freedom is assigned to each cell per field. Also, it overcomes the problem of the failure of Leibniz rule, a common barrier in developing discretized manifolds theory \cite{Liebnitz}.

Beyond the general relativistic context, the notion of curvature is gaining a lot of attention and in particular in the study of networks and simplicial complexes, which are higher-order networks that encapsulate higher-order interactions between nodes, edges, triangles, etc... More recently, great developments have been made in network geometry \cite{mulder2018network, boguna2021network, krioukov2010hyperbolic, barthelemy2011spatial}, which is at the crossroad between network theory and quantum gravity \cite{majid2013noncommutative, gromov1987hyperbolic, bauer2011ollivier, lin2011ricci, lin2010ricci}. The latter depends on a discretization of space-time producing structures, which can be best represented as networks, simplicial complexes, or equivalently higher-order networks \cite{bianconi2021higher, bialas1996focusing, bialas1999phase}.
The interplay between geometry and networks carries over a wide range of applications; for example, hyperbolic spaces with their negative curvatures were shown to be proper embedding latent spaces for a variety of biological and social networks with power-law degree distributions \cite{papadopoulos2012popularity, garcia2016hidden, serrano2008self, albert2014topological, borassi2015hyperbolicity, petri2014homological}. With this approach, we list some applications in routing optimization of the internet and of the trade network \cite{boguna2010sustaining, sulyok2023greedy}. Other examples include network cosmology, which was introduced in the context of relating hyperbolic networks to causal sets \cite{krioukov2012network}. The latter has been explored in the characterization of citation networks' effective dimension \cite{clough2016dimension}. The crossbreeding between the fields also includes causal sets \cite{bombelli1987space, dowker2004quantum} and causal dynamical triangulations \cite{frohlich1992non, loll2019quantum, sornette2012dragon}, quantum network manifolds as models of discrete manifolds relating quantum statistics to emergent network geometry \cite{bialas1999phase}, and in all of these examples curvature is a significant metric in the characterization of their properties.\newline
We thus foresee that our work could have applications in the networks geometry, as it applies to pure $d=2$ simplicial complexes.


\section{Theory: Definition of Curvature}

In this section, we will address the problem of how to calculate the metric and curvature of an arbitrary two-dimensional closed surface. Any two-dimensional surface can be embedded into a three-dimensional Euclidean space and could be
defined by some equation of the form
\begin{equation}
\phi\left(  x,y,z\right)  =0
\label{phieq}
\end{equation}
The metric for such a surface could be obtained from the Euclidean metric
\begin{align*}
ds^{2}  &  = dx^{2}+dy^{2}+dz^{2}.
\end {align*}
Using
\begin{align*}
0  &  =\frac{\partial\phi}{\partial x}dx+\frac{\partial\phi}{\partial y}dy+\frac{\partial\phi}{\partial z}dz
\end{align*}
to eliminate one of the variables, say $z$, we get the curved metric
\begin{align*}
ds^{2}  &  =dx^{2}+dy^{2}+\frac{1}{\left(  \partial_{z}\phi\right)  ^{2}%
}\left(  \partial_{x}\phi\,dx+\partial_{y}\phi\,dy\right)  ^{2}\\
&  =dx^{2}\left(  1+\left(  \frac{\partial_{x}\phi}{\partial_{z}\phi}\right)
^{2}\right)  +dy^{2}\left(  1+\left(  \frac{\partial_{y}\phi}{\partial_{z}%
\phi}\right)  ^{2}\right)  +2dx\,dy\frac{\partial_{x}\phi\partial_{y}\phi
}{\left(  \partial_{z}\phi\right)  ^{2}}%
\end{align*}
with the understanding that equation \ref{phieq} could be solved to express $z$ as a function of $x$ and $y$.\newline

\noindent Given a surface, say that of an apple or a human face, we first consider a Cartesian system, divide the space into a fine cubic lattice with a very small fundamental length scale, and determine the coordinates of all points on the surface to the nearest integer. The next step is to connect every point to its nearest neighbors. Every near three
points are connected to form a triangle. To define what forms the nearest point, we have to calculate distances. 
In differential geometry, the distance between two points on a surface is defined as the geodesic distance. This, however, because of the very small scale of the lattice, is
approximately given by the Euclidean three-dimensional distance
\begin{equation*}
d_{12}^{2}=\left( x_{2} - x_{1} \right)^{2} + \left( y_{2} -y_{1}\right)^{2} + \left( z_{2} - z_{1}\right)^{2}.
\end{equation*}
We will connect every point to three of its nearest neighbors to create a mesh. We require the arrangement to be such that every point is surrounded by four triangles. To calculate the metric to be used on the surface of every triangle, we denote the vertices of one triangle by $A, B, C$ with coordinates
$\left(  x_{a},y_{a},z_{a}\right)  ,$ $\left(  x_{b},y_{b},z_{b}\right)  ,$
$\left(  x_{c},y_{c},z_{c}\right).$ To find the equation of the plane, note that the product $\overrightarrow{AB}\times\overrightarrow{AC}=\overrightarrow
{n}$ is perpendicular to the plane. Thus
\begin{align*}
\overrightarrow{AB} &  =\left(  x_{b}-x_{a}\right)  \overrightarrow{i}+\left(
y_{b}-y_{a}\right)  \overrightarrow{j}+\left(  z_{b}-z_{a}\right)
\overrightarrow{k}\\
\overrightarrow{AC} &  =\left(  x_{c}-x_{a}\right)  \overrightarrow{i}+\left(
y_{c}-y_{a}\right)  \overrightarrow{j}+\left(  z_{c}-z_{a}\right)
\overrightarrow{k}
\end{align*}
\begin{align*}
\overrightarrow{n} &  =\left(  \left(  y_{b}-y_{a}\right)  \left(  z_{c}%
-z_{a}\right)  -\left(  z_{b}-z_{a}\right)  \left(  y_{c}-y_{a}\right)
\right)  \overrightarrow{i}\nonumber\\
&  +\left(  \left(  z_{b}-z_{a}\right)  \left(  x_{c}-x_{a}\right)  -\left(
x_{b}-x_{a}\right)  \left(  z_{c}-z_{a}\right)  \right)  \overrightarrow
{j}\nonumber\\
&  +\left(  \left(  x_{b}-x_{a}\right)  \left(  y_{c}-y_{c}\right)  -\left(
y_{b}-y_{a}\right)  \left(  x_{c}-x_{a}\right)  \right)  \overrightarrow{k}
\end{align*}
\begin{align}\label{normal}
 \overrightarrow{n}  \equiv\alpha_{a}\overrightarrow{i}+\beta_{a}\overrightarrow{j}+\gamma
_{a}\overrightarrow{k}%
\end{align}
The equation of a plane is determined from the condition $\overrightarrow {n}\cdot\overrightarrow{AP}=0$, where the coordinates of $P$ are $\left(x,y,z\right).$ Therefore
\begin{equation*}
0=\alpha_{a}\left(  x-x_{a}\right)  +\beta_{a}\left(  y-y_{a}\right)
+\gamma_{a}\left(  z-z_{a}\right)
\end{equation*}
Solving for $z$ in terms of $x$ and $y$, we get
\begin{equation*}
z=z_{a}-\frac{1}{\gamma_{a}}\left(  \alpha_{a}\left(  x-x_{a}\right)
+\beta_{a}\left(  y-y_{a}\right)  \right)
\end{equation*}
For the metric we use
\begin{equation*}
\Delta z = -\frac{1}{\gamma_{a}}\left(  \alpha_{a}\Delta x+\beta_{a}\Delta y \right)
\end{equation*}
so that
\begin{align*}
\Delta s^{2} &= \Delta x^{2}+\Delta y^{2}+\frac{1}{\gamma_{a}^{2}}\left(\alpha_{a}\Delta x+\beta_{a}\Delta y\right)  ^{2}\\
&  =\left(  1+\frac{\alpha_{a}^{2}}{\gamma_{a}^{2}}\right)  \Delta x^{2}+\left(  1+\frac{\beta_{a}^{2}}{\gamma_{a}^{2}}\right)  \Delta y^{2} + 2\frac{\alpha_{a}\beta_{a}}{\gamma_{a}^{2}}\Delta x\Delta y
\end{align*}
Denote
\begin{equation*}
\overline{\alpha}=\frac{\alpha}{\gamma},\quad\overline{\beta}=\frac{\beta
}{\gamma}
\end{equation*}
then the components of the metric are
\begin{equation*}
g_{\overset{.}{1} \overset{.}{1}}= 1 + \overline{\alpha}^{2}, \quad g_{\overset{.}{2} \overset{.}{2}} = 1 + \overline{\beta}^{2}, \quad g_{\overset{.}{1} \overset{.}{2}} = \overline{\alpha} \overline{\beta}
\end{equation*}
Defining the zweibein, which is related to the metric, by
\begin{equation*}
g_{\mu\nu}=e_{\mu}^{a}e_{\nu}^{a}
\end{equation*}
then $e_{\mu}^{a}$ is not determined uniquely. The simplest choice is
\begin{equation}\label{zweibein}
e_{\overset{,}{1}}^{1}=\sqrt{\frac{1+\overline{\alpha}^{2}+\overline{\beta
}^{2}}{1+\overline{\beta}^{2}}},\quad e_{\overset{,}{1}}^{2}=\frac
{\overline{\alpha}\overline{\beta}}{\sqrt{1+\overline{\beta}^{2}}},\quad
e_{\overset{.}{2}}^{1}=0,\quad e_{\overset{,}{2}}^{2}=\sqrt{1+\overline{\beta
}^{2}}
\end{equation}
The inverse zweibein is
\begin{equation}\label{invzweibein}
e_{1}^{\overset{.}{1}}=\sqrt{\frac{1+\overline{\beta}^{2}}{1+\overline{\alpha
}^{2}+\overline{\beta}^{2}}},\quad e_{1}^{\overset{.}{2}}=-\frac
{\overline{\alpha}\overline{\beta}}{\sqrt{\left(  1+\overline{\beta}%
^{2}\right)  \left(  1+\overline{\alpha}^{2}+\overline{\beta}^{2}\right)  }%
},\quad e_{2}^{\overset{.}{1}}=0,\quad e_{2}^{\overset{.}{2}}=\frac{1}%
{\sqrt{1+\overline{\beta}^{2}}}%
\end{equation}
This procedure is done for every triangle, and then the curvature is calculated using the connection, which is computed using the above zweibein. The connection and the curvature $R_{\mu\nu}\left(  n\right)$ are given in \cite{discretegravity} (refer to the appendix \ref{app} for more details):
\begin{equation*}
\Omega_{\mu}\left(  n\right)  =\cos\frac{1}{2}\mathcal{\ell}^{\mu}\omega_{\mu
}\left(  n\right)  +\tau\sin\frac{1}{2}\mathcal{\ell}^{\mu}\omega_{\mu}\left(
n\right),
\end{equation*}
and 
\begin{equation}\label{rmunu}
R_{\mu\nu}\left(  n\right)  =\frac{1}{\mathcal{\ell}^{\mu}\mathcal{\ell}^{\nu
}}\sin\left(  \frac{\mathcal{\ell}^{\mu}\mathcal{\ell}^{\nu}}{2}\left(
\Delta_{\mu}\omega_{\nu}-\Delta_{\nu}\omega_{\mu}\right)  \right)  \tau,
\end{equation}
where
\begin{equation*}
\Delta_{\overset{.}{1}}\omega_{\overset{.}{2}}=\frac{\omega_{\overset{.}{2}%
}\left(  n+\overset{.}{1}\right)  -\omega_{\overset{.}{2}}\left(  n\right)
}{\mathcal{\ell}^{\overset{.}{1}}},\quad
\Delta_{\overset{.}{2}}\omega_{\overset{.}{1}}=\frac{\omega_{\overset{.}{1}%
}\left(  n+\overset{.}{2}\right)  -\omega_{\overset{.}{1}}\left(  n\right)
}{\mathcal{\ell}^{\overset{.}{2}}},%
\end{equation*}
where $n= (n_1,n_2)$ is a set of two integers, which can be positive or negative, used to enumerate each triangle in space and is associated with the triangle's sides' relative orientations and proximity to the reference triangle's sides, $l^{\mu}$ and $l^{\nu}$ are the lengths of the triangle's sides along the $n_1$ and $n_2$ directions, and finally, $\left(  n+\overset{.}{1}\right)$ and $\left(  n+\overset{.}{2}\right)$ denote the shift in $n_1$ and $n_2$, respectively, for which $\omega_{\overset{.}{2}%
}\left(  n+\overset{.}{1}\right) = \omega_{\overset{.}{2}
}\left(  n_1 +1, n_2 \right)$, and $\omega_{\overset{.}{1}%
}\left( n+\overset{.}{2}\right) = \omega_{\overset{.}{1}
}\left( n_1, n_2 +1 \right)$.\newline

\noindent The torsion condition gives:
\begin{equation*}
0=\frac{1}{\mathcal{\ell}^{\mu}}\left(  \left(  \cos\mathcal{\ell}^{\mu}%
\omega_{\mu}\left(  n\right)  \delta_{ab}+\sin\mathcal{\ell}^{\mu}\omega_{\mu
}\left(  n\right)  \epsilon_{ab}\right)  e_{\nu}^{b}\left(  n+\widehat{\mu
}\right)  -e_{\nu}^{b}\left(  n\right)  \delta_{ab}\right)  \tau_{a}%
-\mu\leftrightarrow\nu.
\end{equation*}
These are two conditions that could be solved to determine the two connections $\omega_{\mu}\left(  n\right).$ By writing these equations explicitly, we get
\begin{align} \label{eq1}
0  & =\frac{1}{\mathcal{\ell}^{\overset{.}{1}}} \left(  \left(  \cos \mathcal{\ell}^{\overset{.}{1}}\omega_{\overset{.}{1}}\left(  n\right) e_{\overset{.}{2}}^{1} \left(  n+ {\overset{.}{1}}\right) + \sin\mathcal{\ell}^{\overset{.}{1}} \omega_{\overset{.}{1}}\left(  n\right)
e_{\overset{.}{2}}^{2}\left(  n+ {\overset{.}{1}}\right)  \right) - e_{\overset{.}{2}}^{1}\left(  n\right)  \right)  \nonumber\\
& -\frac{1}{\mathcal{\ell}^{\overset{.}{2}}}\left(  \left(  \cos\mathcal{\ell}^{\overset{.}{2}} \omega_{\overset{.}{2}}\left(  n\right)  e_{\overset{.}{1}}^{1}\left(  n+ {\overset{.}{2}}\right)  + \sin\mathcal{\ell}^{\overset{.}{2}} \omega_{\overset{.}{2}} \left(  n\right)  e_{\overset{.}{1}}^{2} \left( n+\widehat{\overset{.}{2}}\right)  \right)  - e_{\overset{.}{1}}^{1}\left(  n\right)  \right),
\end{align}
and
\begin{align}\label{eq2}
0  & =\frac{1}{\mathcal{\ell}^{\overset{.}{1}}} \left(  \left(  \cos \mathcal{\ell}^{\overset{.}{1}}\omega_{\overset{.}{1}}\left(  n\right) e_{\overset{.}{2}}^{2} \left( n +{\overset{.}{1}}\right) - \sin\mathcal{\ell}^{\overset{.}{1}} \omega_{\overset{.}{1}}\left(  n\right) e_{\overset{.}{2}}^{1} \left(  n+ {\overset{.}{1}}\right)  \right)
-e_{\overset{.}{2}}^{2}\left(  n\right)  \right)  \nonumber\\
& -\frac{1}{\mathcal{\ell}^{\overset{.}{2}}} \left(  \left(  \cos\mathcal{\ell}^{\overset{.}{2}} \omega_{\overset{.}{2}} \left(  n\right)  e_{\overset{.}{1}}^{2} \left( n + {\overset{.}{2}}\right)  - \sin\mathcal{\ell}^{\overset{.}{2}} \omega_{\overset{.}{2}}\left(  n\right)  e_{\overset{.}{1}}^{1} \left(  n+ {\overset{.}{2}}\right)  \right)  -e_{\overset{.}{1}}^{2}\left(  n\right)  \right),
\end{align}
where $n + {\overset{.}{1}}$ and $ n + {\overset{.}{2}}$, again denote the shift operation in the $n_1$ and $n_2$ indices, respectively. For example, $e_{\overset{.}{2}}^{1}\left(  n+ {\overset{.}{1}}\right)$ corresponds to $e_{\overset{.}{2}}^{1}\left( n_1 + 1, n_2\right)$. \paragraph{}

\section{Methodology}

\begin{figure}[!htp]
    \centering
    \includegraphics[width=0.6\textwidth]{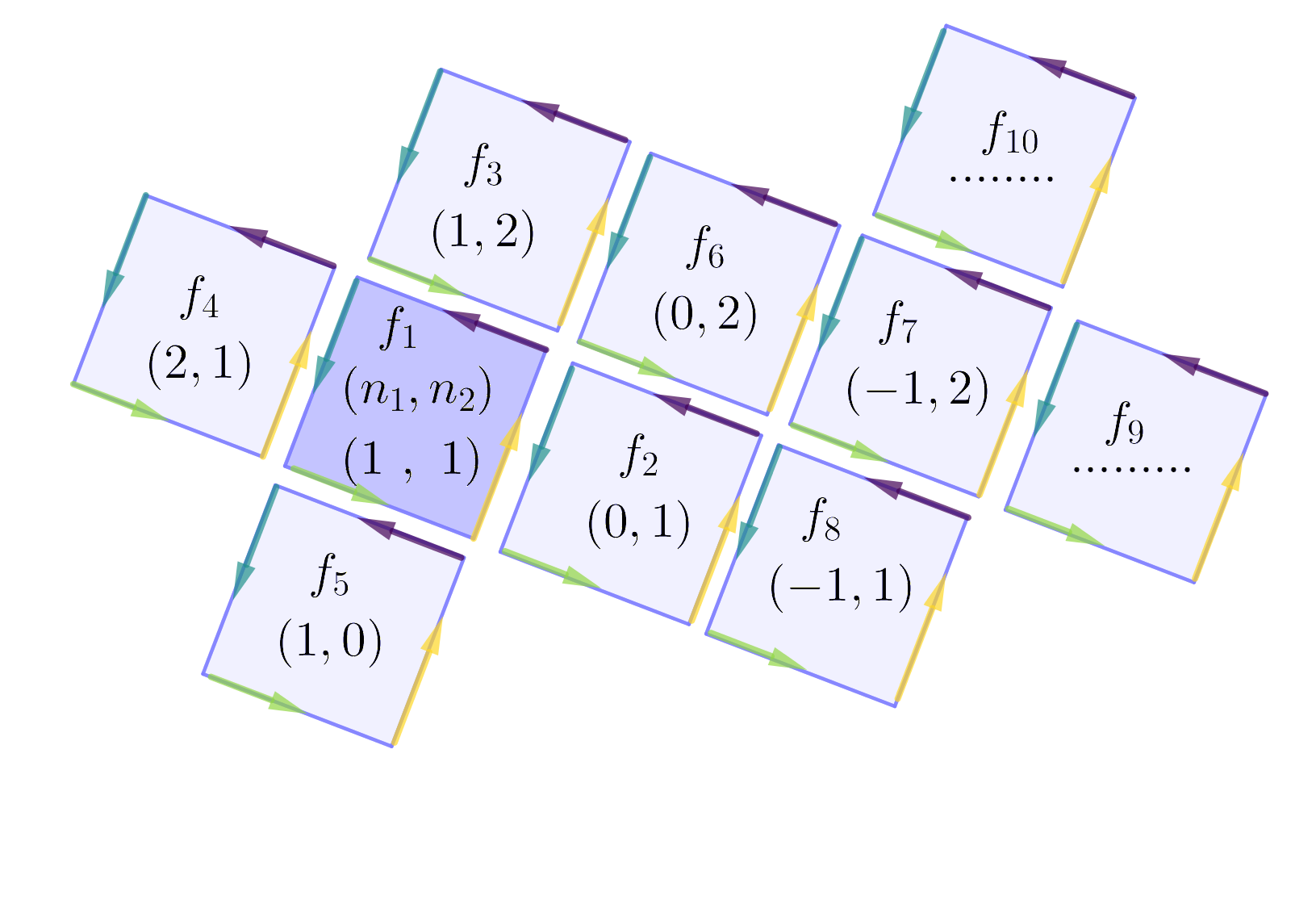}
    \caption{Illustration of the indexing process starting with the seed face $f_1$ is shown.}
    \label{fig:fig1}
\end{figure}

\begin{figure}[!htp]
     \centering
    \includegraphics[width=0.8\textwidth]{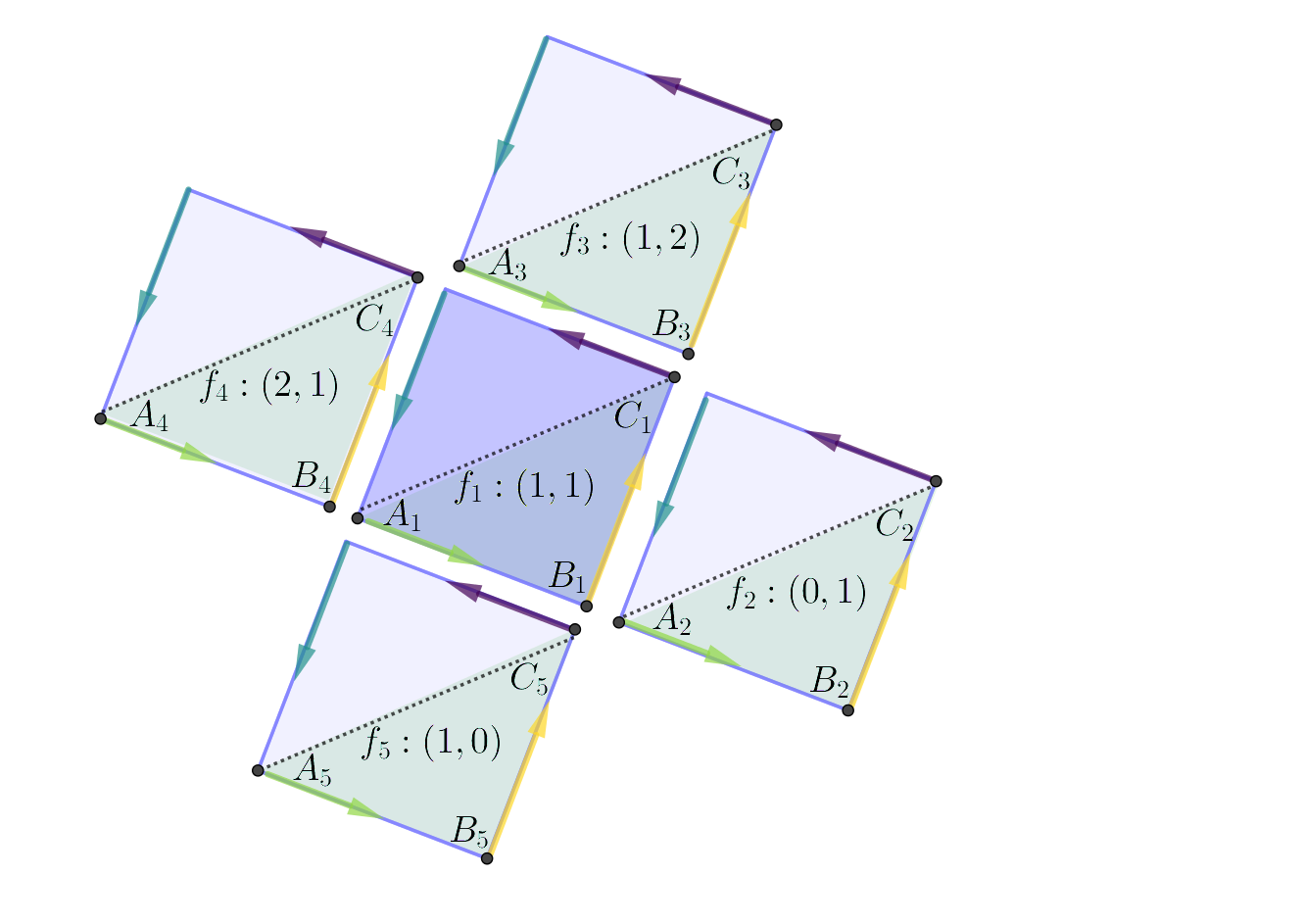}
    \caption{The subdivision of the quads into triangles in order to define the normals to the faces. }
    \label{fig:fig2}
\end{figure}

\begin{figure}[!htp]
    \includegraphics[width=1.1\textwidth]{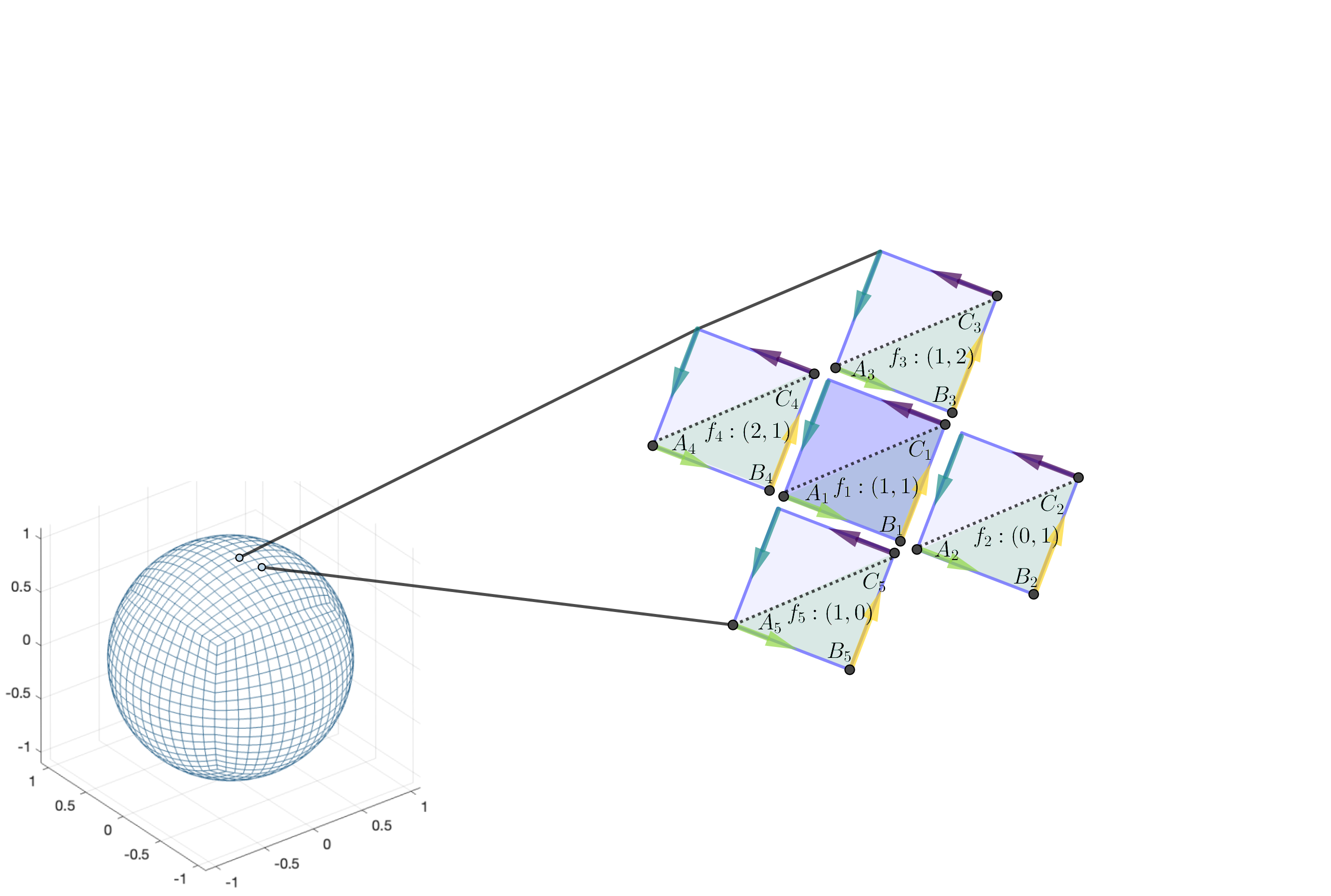}
    \caption{Covering of the whole three-dimensional shape with the the labeled triangles. }
    \label{fig:fig3}
\end{figure}

Given the above derivation, we provide a description of the algorithm that evaluates the discrete curvature of an arbitrary surface. 
Our construction relies on a discretization into quad meshes of the surface of the three-dimensional shape in question. That is every cell should have exactly four neighbours. We start with a seed face, $f_1$ to which we associate $(n_1 = 1, n_2 = 1)$. $f_1$ is then given an orientation either anticlockwise as shown in Figure \ref{fig:fig1} (or clockwise), which subsequently define the directions of increase/decrease of $n_1$ and $n_2$.
In our illustrative example in Figure \ref{fig:fig1}, the yellow vector defines the direction of increase in $n_1$, the purple vector the direction of decrease in $n_2$. Further, since we do not require the quads to be squares or rectangles, but rather four-sided polygons, it is then clear that the side that does not share any common vertex with the yellow one is defined as the direction of decrease in $n_1$, and that which does not have any common vertex with the purple vector is defined to be the direction of increase in $n_2$.
More explicitly, let's define the indices of $f_2$, $f_3$, $f_4$, and $f_5$. The side that is common between $f_1$ and $f_2$ is in the opposite direction to the yellow vector in $f_1$, and thus $n_1$ of $f_2$ is that of $f_1$ minus 1, while $n_2$ of $f_2$ is unchanged since there is no intersection along the purple side. The side in $f_4$ that intersects with $f_1$ is parallel to the yellow side in $f_1$ and thus $n_1$ of $f_4$ is 2. Similarly for the other direction, the side in $f_3$ that is common with $f_1$ is anti-parallel to the purple and thus $n_2$ of $f_3$ is 2 while that of $f_5$ is zero. 
The process is done iteratively from the seed node to its neighbors which creates an ordered list of faces that covers all of the surface of the three-dimensional shape.  We note that the resultant structure is a network of average degree $k=4$ and a corresponding uniform degree distribution, where the degree denotes the number of links for each vertex or point in space \cite{newman2018networks}. 
\newpage
After the indexing is finished, we move on to define the lower triangle of the quad as our basic unit as shown in Figure \ref{fig:fig2}, since we need it to calculate the normal to the face, given in Equation \ref{normal}; the lower triangle inherits the indices of its circumscribing quad. Finally the full three-dimensional shape is covered by these labeled triangles as show in Figure \ref{fig:fig3}.  These triangles are simplices of degree $d=2$ and form a $d=2$ pure simplicial complex, since all its faces are triangles \cite{bianconi2021higher}. A simplex characterizes an interaction between two or more nodes, for example a $0-$simplex is a node, a 1-simplex is a an edge, and $2$ simplex is a triangle.   We can now compute the 
zweibeins and inverse zweibein given in Equations \ref{zweibein} and \ref{invzweibein}. The latter can be finally plugged into Equations \ref{eq1} and \ref{eq2} to solve for the spin connections. 

\newpage
Now Equations \ref{eq1} and \ref{eq2} are solved in {\sf R} using {\sf nleqslv} package. The solution of the latter is then plugged into Equation \ref{rmunu}, which yields the curvature tensor defined for every triangular face. It is worth noting that our definition of curvature applies to pure $d=2$ simplicial complexes. 
The github repository associated with this paper is available at the following 
\href{https://github.com/s-najem/Surface-Orientation-and-Scalar-Curvature/tree/main}{link.}
\section{Results}
\begin{figure}[!htp]
    \includegraphics[width=1.1\textwidth]{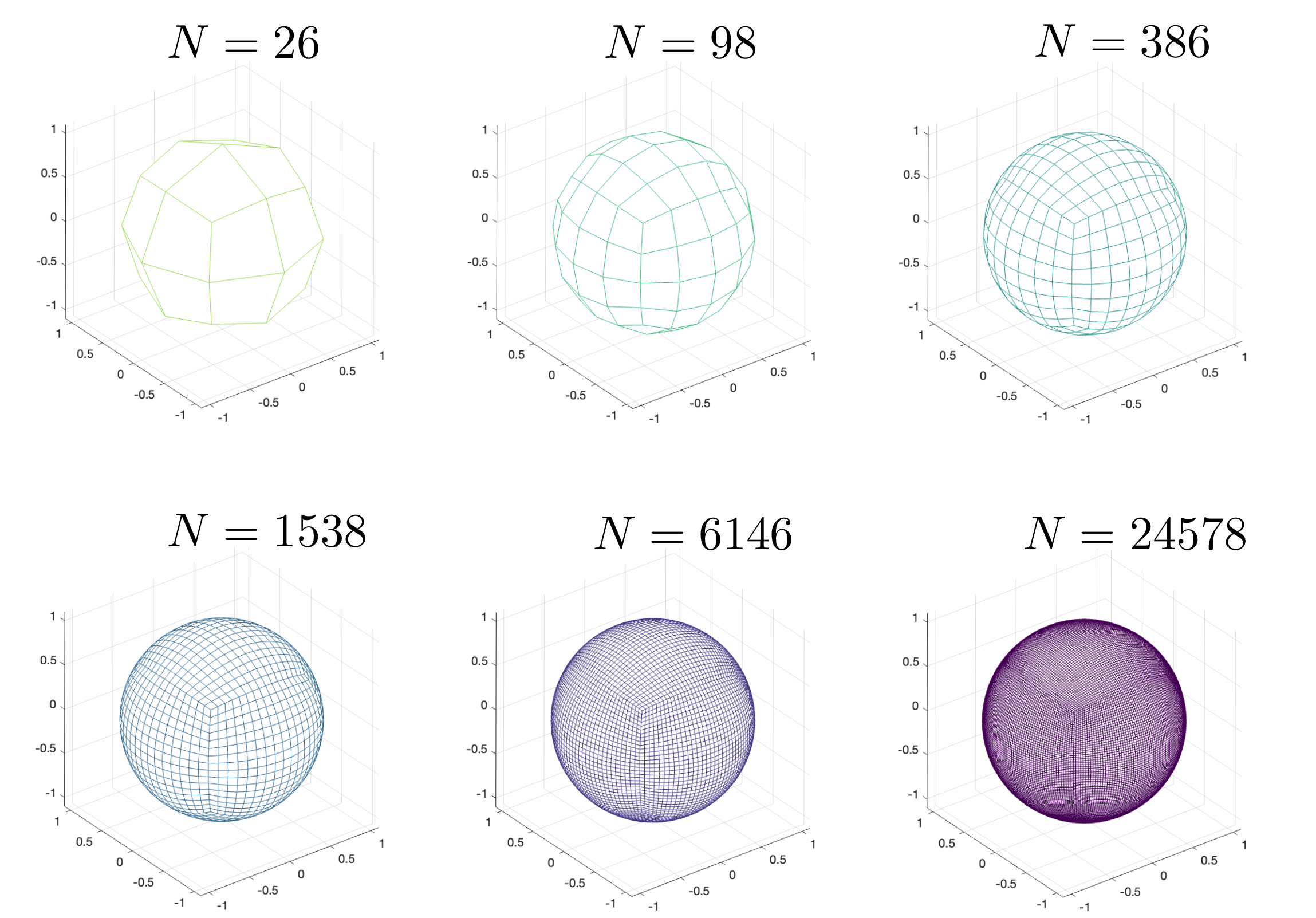}
    \caption{Different discretizations of the sphere with increasing number of cell sizes. }
    \label{fig:fig4}
\end{figure}

\begin{figure}[!htp]
     \centering
    \includegraphics[width=0.5\textwidth]{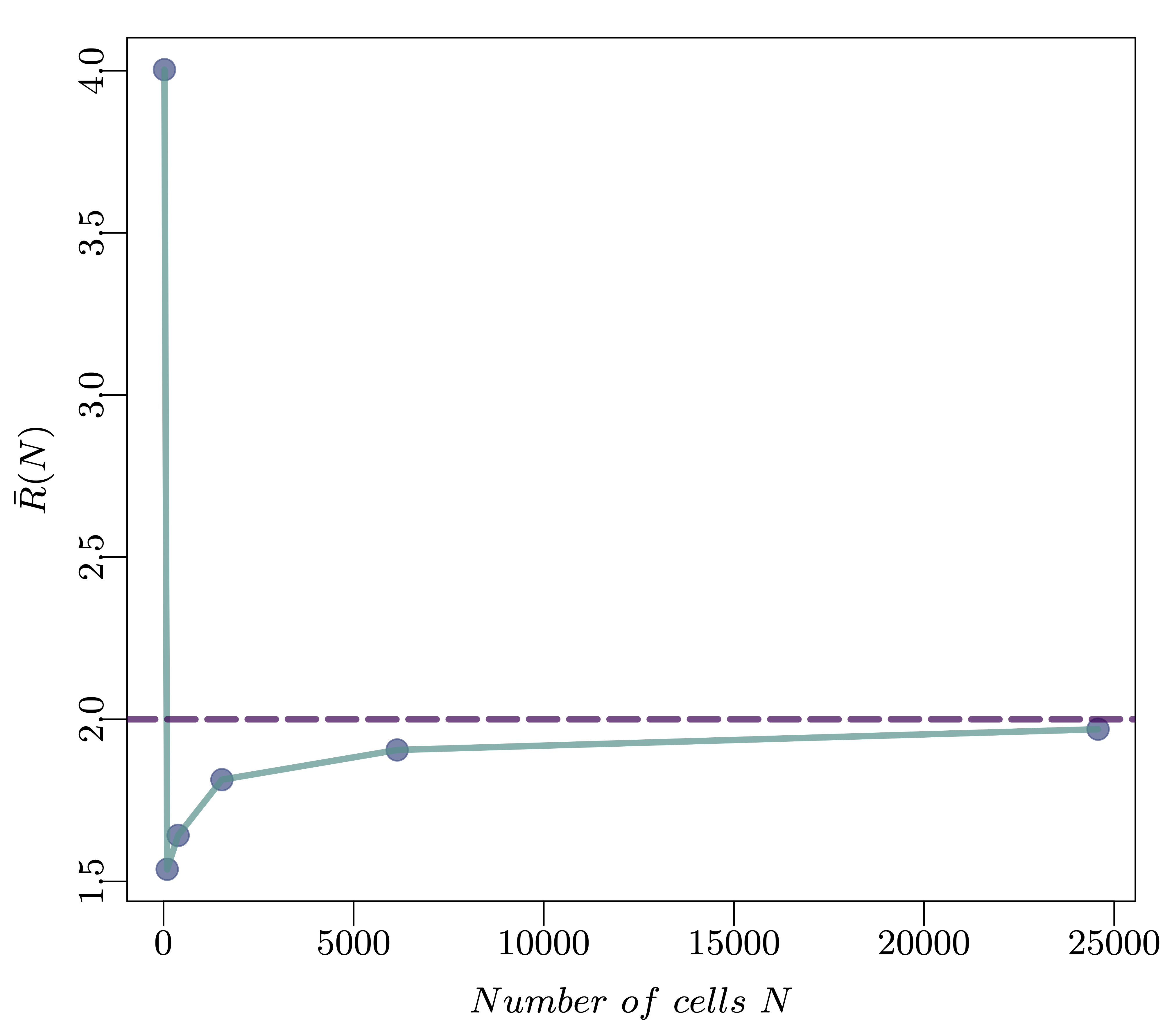}
    \caption{The mean curvature is followed as a function of the number of cells $N$.}
    \label{fig:fig5}
\end{figure}

In this section, we discretize the unit sphere of radius $r=1$ with increasing number of faces $N$ and decreasing cell sizes, as shown in Figure \ref{fig:fig4}, using the { \sf $S^2$ Sampling Toolbox} in {\sf Matlab} \cite{githubGitHubAntonSemechkoS2SamplingToolbox}. Then we evaluate the mean curvature and assess the error associated with each discretization by measuring how far it is from the expected value of $2/r^2 = 2$, as shown in Figure \ref{fig:fig5}. We note that the convergence to the expected value happens at around $N= 24578$.
 
\begin{figure}[!htp]
    \includegraphics[width=0.8\textwidth]{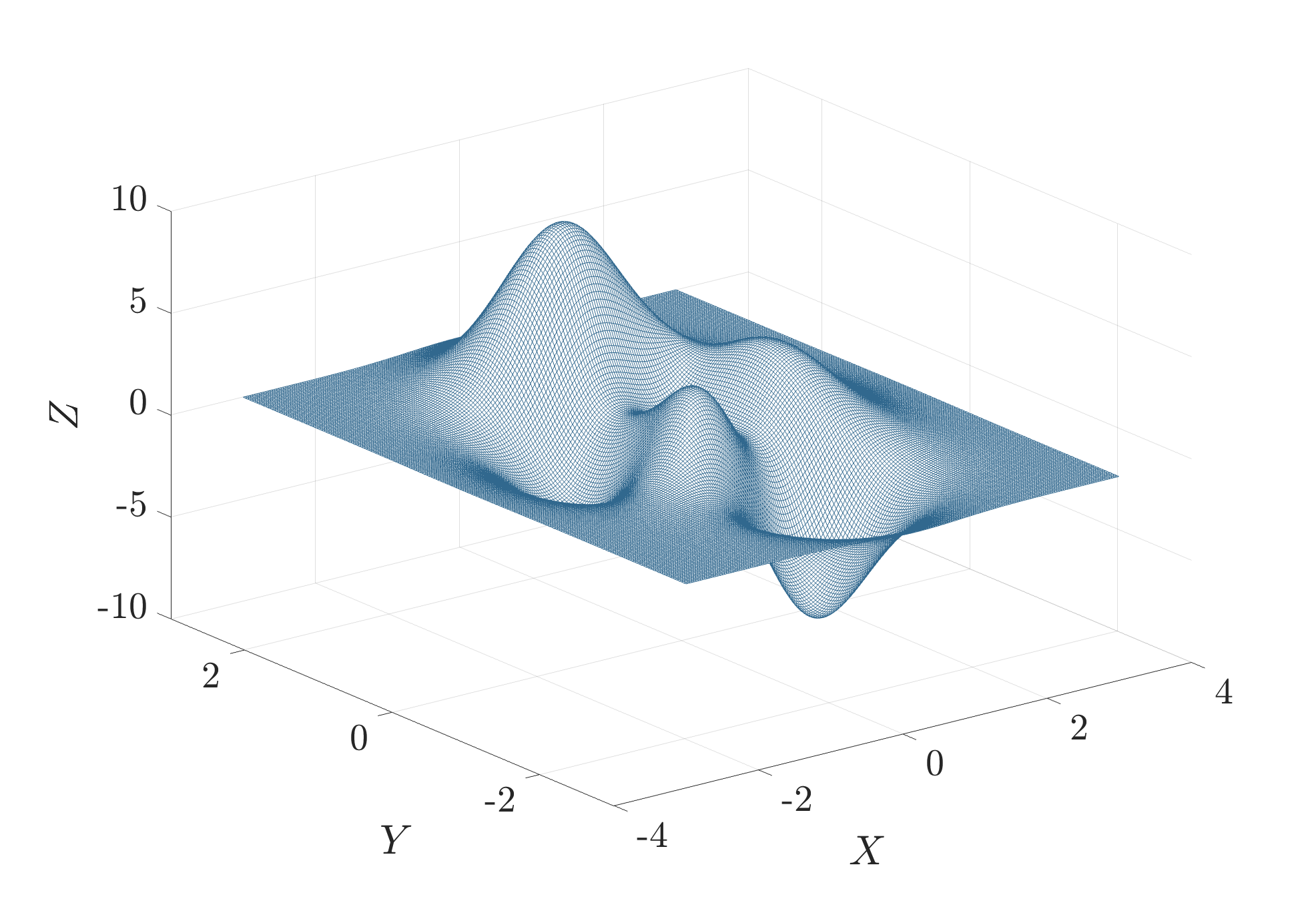}
    \caption{ A three-dimensional plot of the surface in question. }
    \label{fig:fig6}
\end{figure}

\begin{figure}[!htp]
    \includegraphics[width=0.7\textwidth]{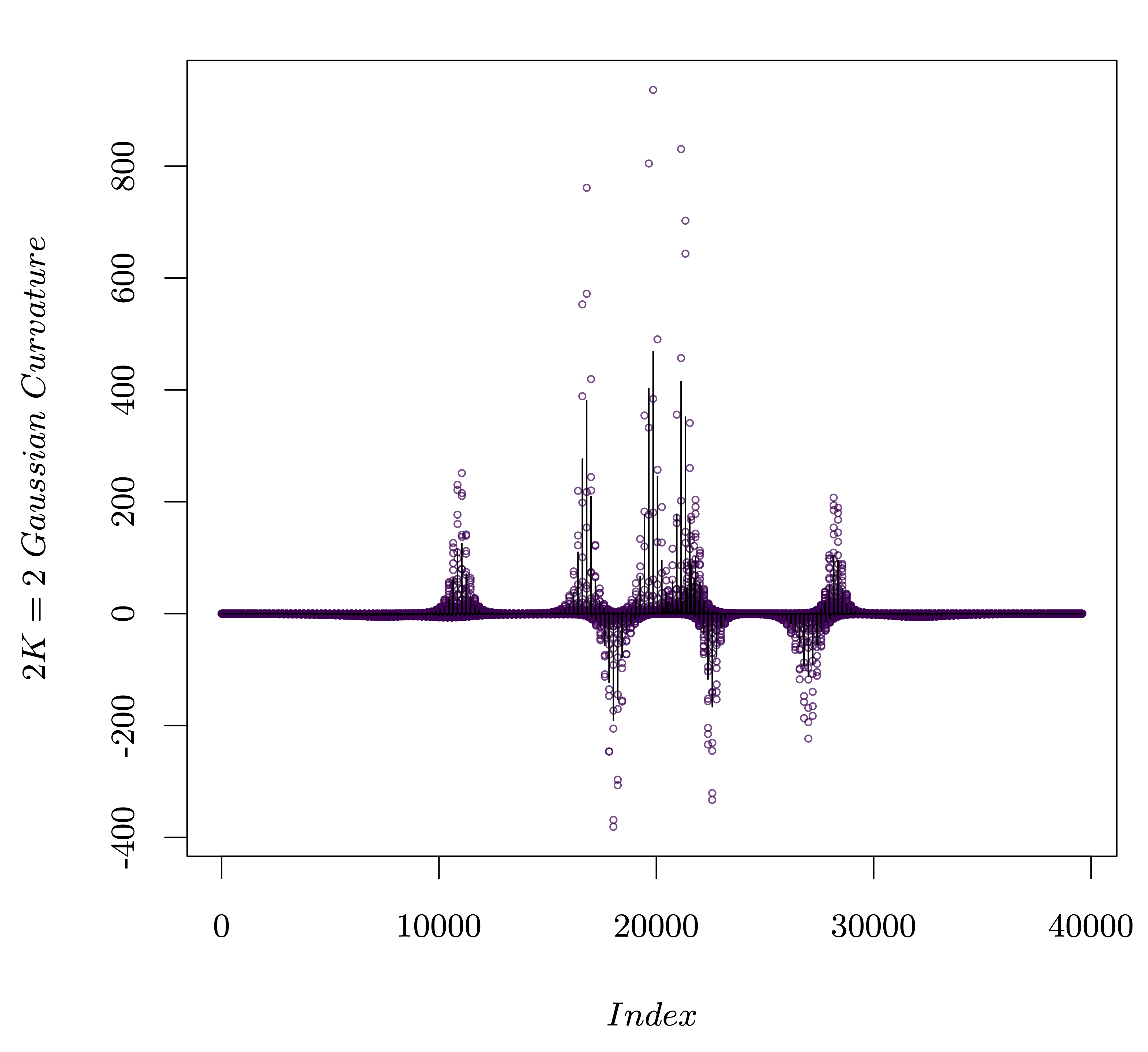}
    \caption{ The analytical value of Gaussian curvature is shown for $N = 40000$ in the figure. }
    \label{fig:fig7}
\end{figure}

\begin{figure}[!htp]
    \includegraphics[width=0.7\textwidth]{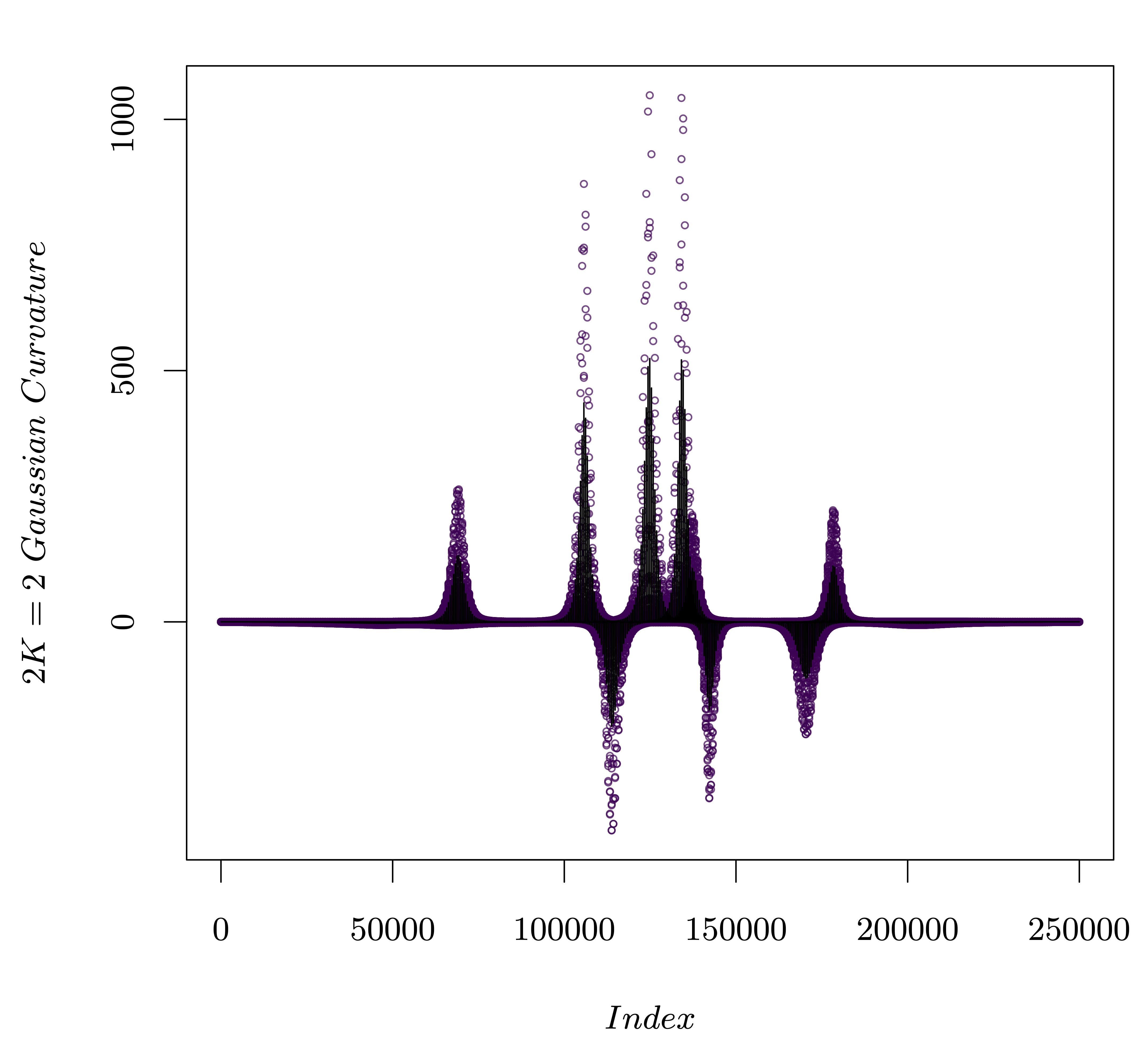}
    \caption{ The analytical value of Gaussian curvature  for $N= 250000$ is shown in the figure. }
    \label{fig:fig70}
\end{figure}

\begin{figure}[!htp]
    \includegraphics[width=0.7\textwidth]{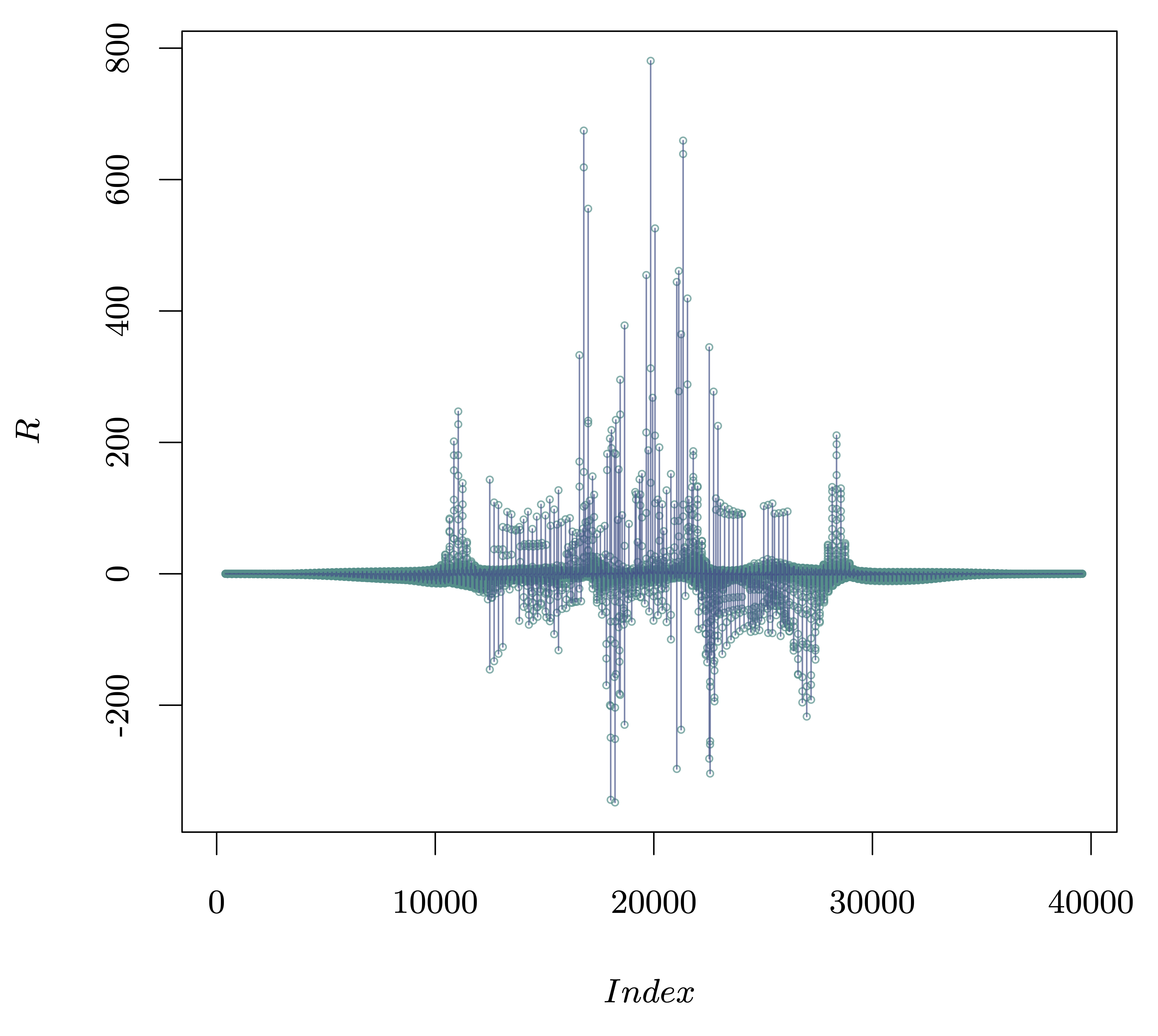}
    \caption{The curvature of the surface using our method is shown with a root-mean-square of  0.085 for $N=40000$. }
    \label{fig:fig8}
\end{figure}

\begin{figure}[!htp]
    \includegraphics[width=0.7\textwidth]{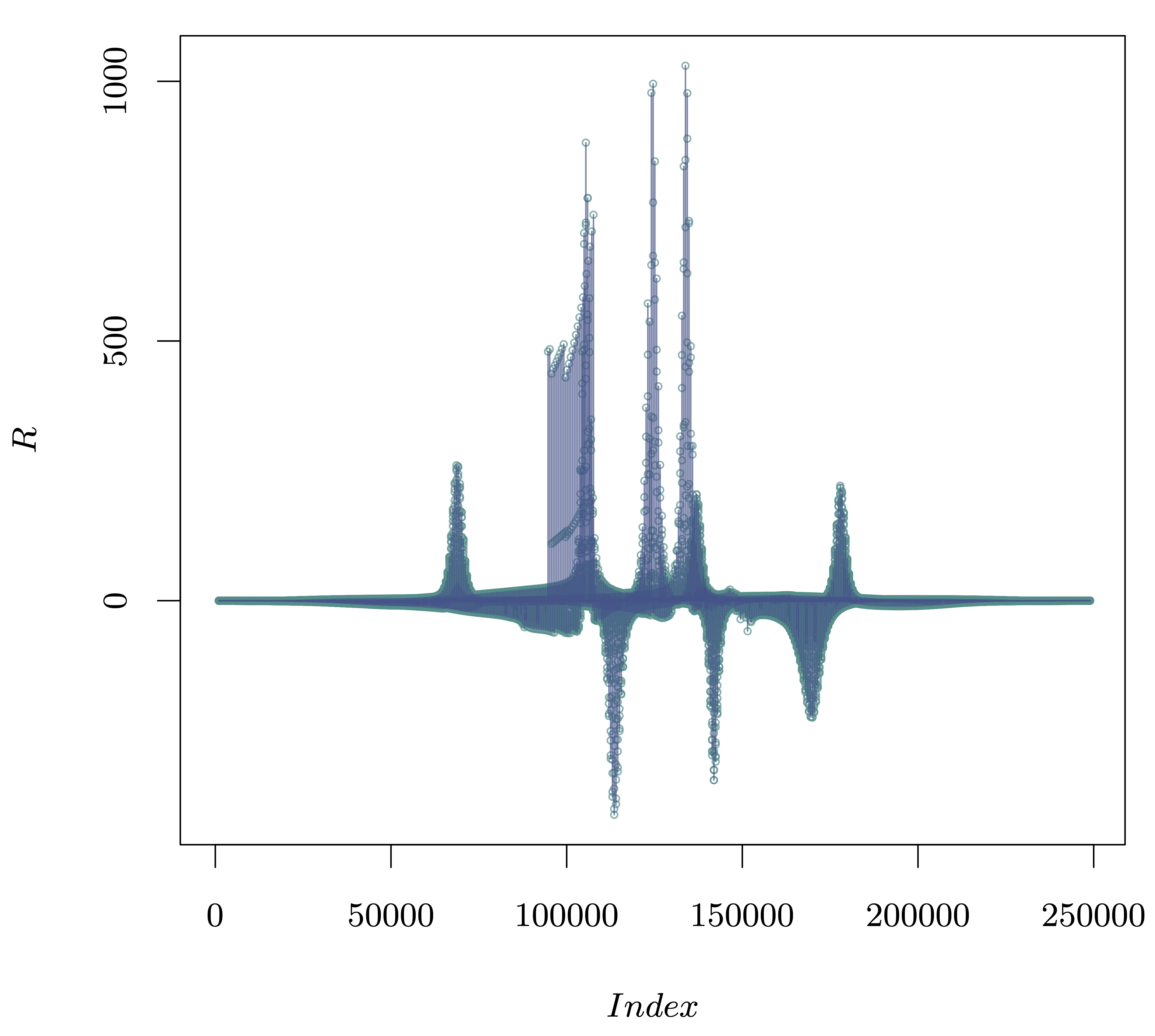}
    \caption{The curvature of the surface for $N= 250000$ using our method is shown with a root-mean-square error of 0.037.}
    \label{fig:fig80}
\end{figure}

\newpage
We then move on to evaluate our scheme on a surface given by: 
\begin{equation*}
Z =  3(1-X)^2e^{-X^2 - (Y+1)^2} - 10(X/5 - X^3 - Y^5)e^{-X^2-Y^2} - 1/3e^{-(X+1)^2 - Y^2}
\label{eqz}
\end{equation*}
shown in Figure \ref{fig:fig6}. In Figures \ref{fig:fig7} and \ref{fig:fig70}, we show the expected analytical value of the Gaussian curvature $R$ for $N = 40000$ and $N=250000$ respectively, as given by {\sf Matlab's Surfature} \cite{mathworksSurfaceCurvature}. We apply our method and show the recovered curvatures in Figures \ref{fig:fig8} and \ref{fig:fig80} again for $N = 40000$ and $N=250000$ respectively. We report a better convergence with a root-mean-square error of 0.037 for the finer discretization compared to 0.085 of the coarser.


\begin{figure}[!htp]
    \includegraphics[width=1.\textwidth]{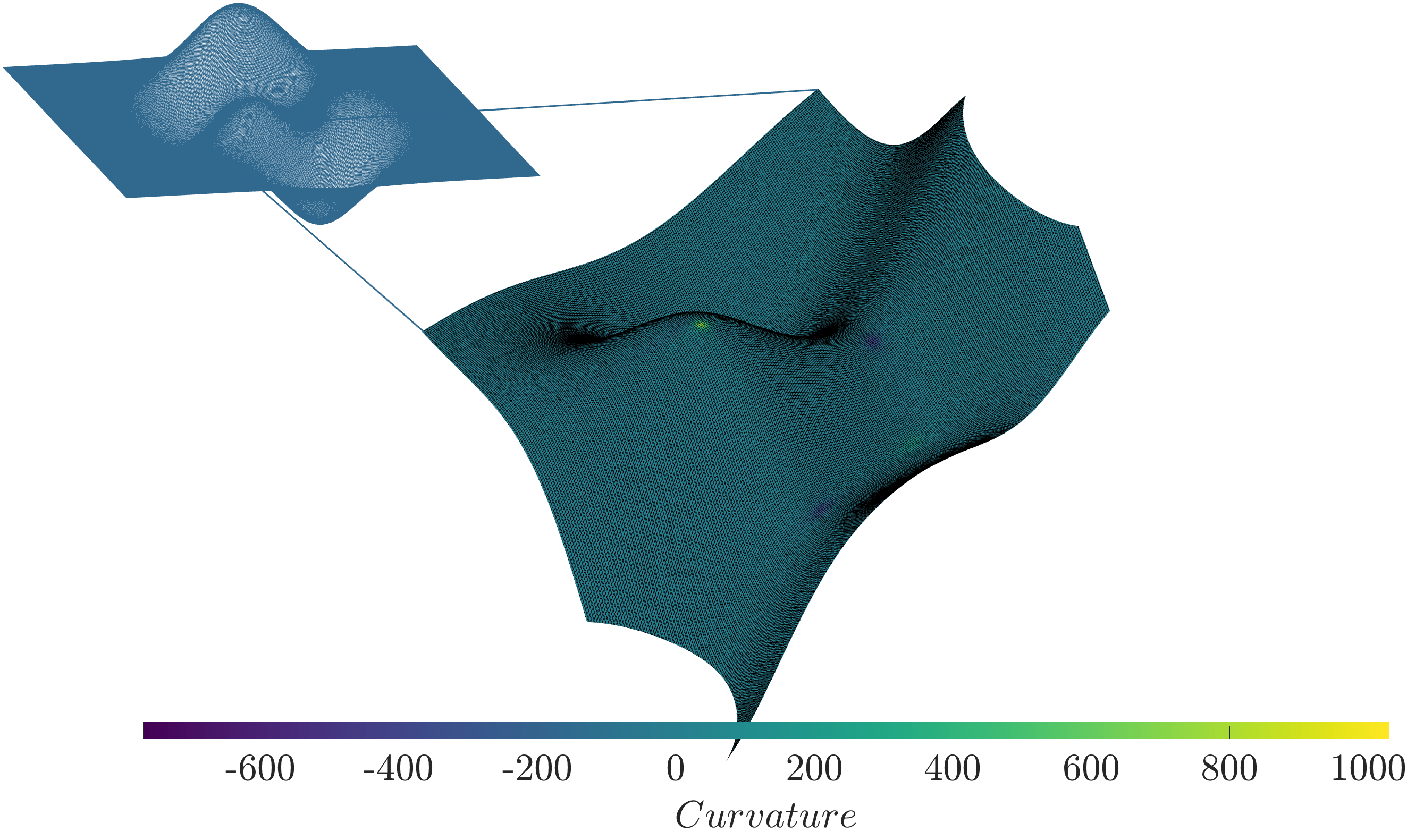}
    \caption{A visualization of a surface plot color coded by the value of the curvature. }
    \label{fig:fig9}
\end{figure}

\begin{figure}[!htp]
    \centering
    \includegraphics[width=0.8\textwidth]{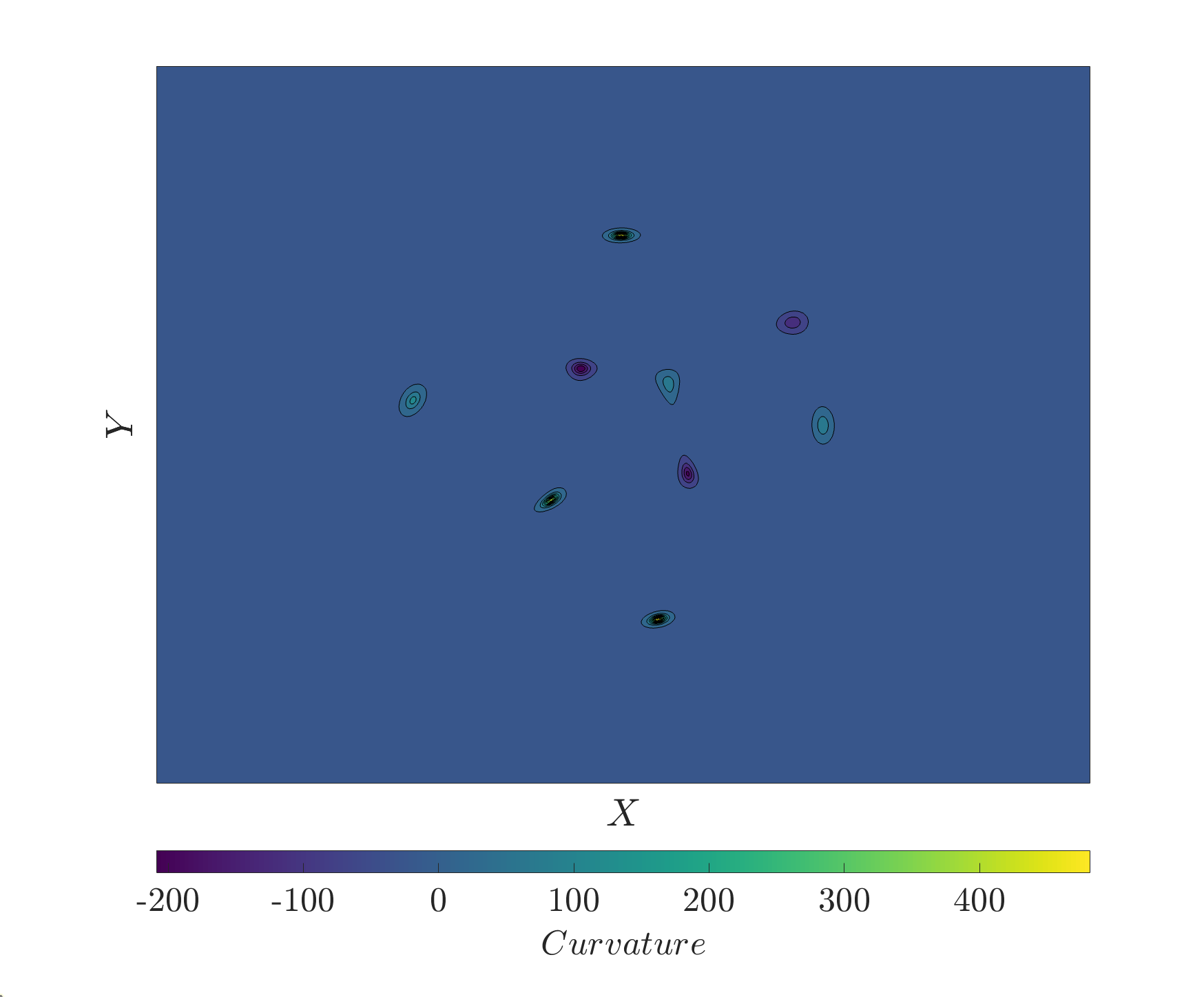}
    \caption{A level plot projection onto the $XY$ plane for $N = 40000$.}
    \label{fig:fig10}
\end{figure}

\newpage
We finally present a visualization of the curvature overlayed on the three-dimensional curve and colorcoded accordingly along with a surface plot projected in the $XY$ plane shown in Figures \ref{fig:fig9} and \ref{fig:fig10} respectively. 


\newpage
\section{Conclusion}

In this paper, we have proposed a method for calculating the curvatures of arbitrary two-dimensional surfaces. We have provided a numerical implementation of the procedure that applies to pure d = 2 simplicial complexes. We first discretized the unit-radius sphere using Matlab's $S^2$ Sampling Toolbox. The mean curvature was calculated and it turned out that the convergence to the expected value happens when the number of faces is around $24578$. Subsequently, this method was also tested on another surface (equation \ref{eqz}), and for a fine discretization of $N=250000$, the calculated value of R converges to the analytical value, as given by Matlab's Surfature, with an error of rms of 0.037. In future work, we will explore the curvature of various arbitrary two-dimensional surfaces.

\section*{Appendix}
\label{app}

The developed theory of discrete manifolds considered in this paper views the space as consisting of elementary cells with no internal differentiable structure. Each elementary cell is completely characterized by displacement operators connecting it to neighboring cells and by the spin connection. The gauge symmetry of the local rotation group and the elements of the spin connection group were used to form the curvature. Parallel transport was defined and spin connections, torsion, and curvature were expressed in terms of the soldering forms in neighboring cells. \newline

\noindent In this appendix, we give the expressions of the connection, the curvature, and the condition of no-torsion, as derived in \cite{discretegravity}. 
Upon considering the shift from the cell $n + 1_\beta$ to the $n-$cell, the parallel transported veilbein is defined using the local spin connection group as
\begin{equation*}
v_{a}^{p.t.} \left(n + 1_{\beta} \rightarrow n \right) = \left( \Omega_{\beta}^{-1}\left(  n\right) \right)_a^b v_b(n),
\end{equation*}
where $\Omega_{\beta}^{-1}\left(  n\right)_a^b$ is the inverse of 
\begin{equation*}
\Omega_{\mu}\left(  n\right) = exp \left( \omega_{\mu}^{cd} (n) J_{cd} \right).
\end{equation*}

\noindent In the continuous limit, $\omega_{\mu}^{cd} (n)$ is the well-known spin connection. $J_{cd}$ are the generators of the rotation group, satisfying the following commutation relation:
\begin{equation*}
    [J_{ab}, J_{cd}] = \frac{1}{2} \left( \delta_{bc} J_{ad} + \delta_{ad} J_{bc} - \delta_{ac} J_{bd} - \delta_{bd} J_{ac} \right) 
\end{equation*}
\noindent The condition of no-torsion takes the following form:
\begin{equation*}
    \left(\Omega_{\beta} \left( n \right) \right)_a^b \ e_{b \alpha} \left( n + 1_{\beta} \right) - e_{a \alpha} = \left(\Omega_{\alpha} \left( n \right) \right)_a^b \ e_{b \beta} \left( n + 1_{\alpha} \right) - e_{a \beta},
\end{equation*}
and the curvature is given by 
\begin{equation*}
    R_{\alpha \beta} (n) = \frac{1}{2} \left( \Omega_{\alpha} (n)\Omega_{\beta} (n + 1_{\alpha}) \Omega^{-1}_{\alpha} (n + 1_{\beta}) \Omega^{-1}_{\beta} (n) - (\alpha \leftrightarrow \beta) \right)
\end{equation*}
where $n + 1_{\beta}$ shows which of the arguments in $n$ is shifted by unity. The elementary cells in 2-dimensional space are enumerated by a $(n_1, n_2)$ set. \newline

\section*{Acknowledgments}
The work of A. H. C is supported in part by the National Science Foundation Grant No. Phys-2207663.

\section*{Supporting Material}
The github repository associated with this paper is available at the following 
\href{https://github.com/s-najem/Surface-Orientation-and-Scalar-Curvature/tree/main}{link.}
\bibliography{main}

\begin{thebibliography}{39}
\providecommand{\natexlab}[1]{#1}
\providecommand{\url}[1]{\texttt{#1}}
\expandafter\ifx\csname urlstyle\endcsname\relax
  \providecommand{\doi}[1]{doi: #1}\else
  \providecommand{\doi}{doi: \begingroup \urlstyle{rm}\Url}\fi

\bibitem[Bauer et~al.(2011)Bauer, Jost, and Liu]{bauer2011ollivier}
Frank Bauer, J{\"u}rgen Jost, and Shiping Liu.
\newblock Ollivier-ricci curvature and the spectrum of the normalized graph
  laplace operator.
\newblock \emph{arXiv preprint arXiv:1105.3803}, 2011.

\bibitem[Ollivier(2007)]{ollivier2007ricci}
Yann Ollivier.
\newblock Ricci curvature of metric spaces.
\newblock \emph{Comptes Rendus Mathematique}, 345\penalty0 (11):\penalty0
  643--646, 2007.

\bibitem[Jost and Liu(2014)]{jost2014ollivier}
J{\"u}rgen Jost and Shiping Liu.
\newblock Ollivier’s ricci curvature, local clustering and
  curvature-dimension inequalities on graphs.
\newblock \emph{Discrete \& Computational Geometry}, 51\penalty0 (2):\penalty0
  300--322, 2014.

\bibitem[Sreejith et~al.(2016)Sreejith, Mohanraj, Jost, Saucan, and
  Samal]{sreejith2016forman}
RP~Sreejith, Karthikeyan Mohanraj, J{\"u}rgen Jost, Emil Saucan, and Areejit
  Samal.
\newblock Forman curvature for complex networks.
\newblock \emph{Journal of Statistical Mechanics: Theory and Experiment},
  2016\penalty0 (6):\penalty0 063206, 2016.

\bibitem[Regge(1961)]{regge1961general}
Tullio Regge.
\newblock General relativity without coordinates.
\newblock \emph{Il Nuovo Cimento (1955-1965)}, 19:\penalty0 558--571, 1961.

\bibitem[Bianconi(2021)]{bianconi2021higher}
Ginestra Bianconi.
\newblock \emph{Higher-order networks}.
\newblock Cambridge University Press, 2021.

\bibitem[Chamseddine and Mukhanov(2021)]{discretegravity}
Ali~H Chamseddine and Viatcheslav Mukhanov.
\newblock Discrete gravity.
\newblock \emph{Journal of High Energy Physics}, 2021\penalty0 (11):\penalty0
  1--13, 2021.

\bibitem[Chamseddine et~al.(2022)Chamseddine, Malaeb, and Najem]{discrete1}
Ali~H Chamseddine, Ola Malaeb, and Sara Najem.
\newblock Scalar curvature in discrete gravity.
\newblock \emph{The European Physical Journal C}, 82\penalty0 (7):\penalty0
  651, 2022.

\bibitem[Chamseddine et~al.(2023)Chamseddine, Malaeb, and Najem]{discrete2}
Ali~H Chamseddine, Ola Malaeb, and Sara Najem.
\newblock Curvature tensor in discrete gravity.
\newblock \emph{The European Physical Journal C}, 83\penalty0 (10):\penalty0
  896, 2023.

\bibitem[Chamseddine et~al.(2024)Chamseddine, Malaeb, and Najem]{blackhole}
Ali~H Chamseddine, Ola Malaeb, and Sara Najem.
\newblock Black hole in discrete gravity.
\newblock \emph{The European Physical Journal C}, 84:\penalty0 284, 2024.

\bibitem[Miller(1964)]{Liebnitz}
K.~Miller.
\newblock \emph{The calculus of finite differences and difference equations}.
\newblock Dover Publications Inc., New York, NY., 1964.

\bibitem[Mulder and Bianconi(2018)]{mulder2018network}
Daan Mulder and Ginestra Bianconi.
\newblock Network geometry and complexity.
\newblock \emph{Journal of Statistical Physics}, 173\penalty0 (3):\penalty0
  783--805, 2018.

\bibitem[Boguna et~al.(2021)Boguna, Bonamassa, De~Domenico, Havlin, Krioukov,
  and Serrano]{boguna2021network}
Marian Boguna, Ivan Bonamassa, Manlio De~Domenico, Shlomo Havlin, Dmitri
  Krioukov, and M~{\'A}ngeles Serrano.
\newblock Network geometry.
\newblock \emph{Nature Reviews Physics}, 3\penalty0 (2):\penalty0 114--135,
  2021.

\bibitem[Krioukov et~al.(2010)Krioukov, Papadopoulos, Kitsak, Vahdat, and
  Bogun{\'a}]{krioukov2010hyperbolic}
Dmitri Krioukov, Fragkiskos Papadopoulos, Maksim Kitsak, Amin Vahdat, and
  Mari{\'a}n Bogun{\'a}.
\newblock Hyperbolic geometry of complex networks.
\newblock \emph{Physical Review E—Statistical, Nonlinear, and Soft Matter
  Physics}, 82\penalty0 (3):\penalty0 036106, 2010.

\bibitem[Barth{\'e}lemy(2011)]{barthelemy2011spatial}
Marc Barth{\'e}lemy.
\newblock Spatial networks.
\newblock \emph{Physics reports}, 499\penalty0 (1-3):\penalty0 1--101, 2011.

\bibitem[Majid(2013)]{majid2013noncommutative}
Shahn Majid.
\newblock Noncommutative riemannian geometry on graphs.
\newblock \emph{Journal of Geometry and Physics}, 69:\penalty0 74--93, 2013.

\bibitem[Gromov(1987)]{gromov1987hyperbolic}
Mikhael Gromov.
\newblock Hyperbolic groups.
\newblock In \emph{Essays in group theory}, pages 75--263. Springer, 1987.

\bibitem[Lin et~al.(2011)Lin, Lu, and Yau]{lin2011ricci}
Yong Lin, Linyuan Lu, and Shing-Tung Yau.
\newblock Ricci curvature of graphs.
\newblock \emph{Tohoku Mathematical Journal, Second Series}, 63\penalty0
  (4):\penalty0 605--627, 2011.

\bibitem[Lin and Yau(2010)]{lin2010ricci}
Yong Lin and Shing-Tung Yau.
\newblock Ricci curvature and eigenvalue estimate on locally finite graphs.
\newblock \emph{Mathematical research letters}, 17\penalty0 (2):\penalty0
  343--356, 2010.

\bibitem[Bialas et~al.(1996)Bialas, Burda, Krzywicki, and
  Petersson]{bialas1996focusing}
P~Bialas, Zdzis{\l}aw Burda, A~Krzywicki, and Bengt Petersson.
\newblock Focusing on the fixed point of 4d simplicial gravity.
\newblock \emph{Nuclear Physics B}, 472\penalty0 (1-2):\penalty0 293--308,
  1996.

\bibitem[Bialas et~al.(1999)Bialas, Burda, and Johnston]{bialas1999phase}
P~Bialas, Zdzis{\l}aw Burda, and D~Johnston.
\newblock Phase diagram of the mean field model of simplicial gravity.
\newblock \emph{Nuclear Physics B}, 542\penalty0 (1-2):\penalty0 413--424,
  1999.

\bibitem[Papadopoulos et~al.(2012)Papadopoulos, Kitsak, Serrano, Bogun{\'a},
  and Krioukov]{papadopoulos2012popularity}
Fragkiskos Papadopoulos, Maksim Kitsak, M~{\'A}ngeles Serrano, Mari{\'a}n
  Bogun{\'a}, and Dmitri Krioukov.
\newblock Popularity versus similarity in growing networks.
\newblock \emph{Nature}, 489\penalty0 (7417):\penalty0 537--540, 2012.

\bibitem[Garc{\'\i}a-P{\'e}rez et~al.(2016)Garc{\'\i}a-P{\'e}rez,
  Bogu{\~n}{\'a}, Allard, and Serrano]{garcia2016hidden}
Guillermo Garc{\'\i}a-P{\'e}rez, Mari{\'a}n Bogu{\~n}{\'a}, Antoine Allard, and
  M{\'A}2016NatSR Serrano.
\newblock The hidden hyperbolic geometry of international trade: World trade
  atlas 1870--2013.
\newblock \emph{Scientific reports}, 6\penalty0 (1):\penalty0 1--10, 2016.

\bibitem[Serrano et~al.(2008)Serrano, Krioukov, and
  Bogun{\'a}]{serrano2008self}
M~{\'A}ngeles Serrano, Dmitri Krioukov, and Mari{\'a}n Bogun{\'a}.
\newblock Self-similarity of complex networks and hidden metric spaces.
\newblock \emph{Physical review letters}, 100\penalty0 (7):\penalty0 078701,
  2008.

\bibitem[Albert et~al.(2014)Albert, DasGupta, and
  Mobasheri]{albert2014topological}
R{\'e}ka Albert, Bhaskar DasGupta, and Nasim Mobasheri.
\newblock Topological implications of negative curvature for biological and
  social networks.
\newblock \emph{Physical Review E}, 89\penalty0 (3):\penalty0 032811, 2014.

\bibitem[Borassi et~al.(2015)Borassi, Chessa, and
  Caldarelli]{borassi2015hyperbolicity}
Michele Borassi, Alessandro Chessa, and Guido Caldarelli.
\newblock Hyperbolicity measures democracy in real-world networks.
\newblock \emph{Physical Review E}, 92\penalty0 (3):\penalty0 032812, 2015.

\bibitem[Petri et~al.(2014)Petri, Expert, Turkheimer, Carhart-Harris, Nutt,
  Hellyer, and Vaccarino]{petri2014homological}
Giovanni Petri, Paul Expert, Federico Turkheimer, Robin Carhart-Harris, David
  Nutt, Peter~J Hellyer, and Francesco Vaccarino.
\newblock Homological scaffolds of brain functional networks.
\newblock \emph{Journal of The Royal Society Interface}, 11\penalty0
  (101):\penalty0 20140873, 2014.

\bibitem[Bogun{\'a} et~al.(2010)Bogun{\'a}, Papadopoulos, and
  Krioukov]{boguna2010sustaining}
Mari{\'a}n Bogun{\'a}, Fragkiskos Papadopoulos, and Dmitri Krioukov.
\newblock Sustaining the internet with hyperbolic mapping.
\newblock \emph{Nature communications}, 1\penalty0 (1):\penalty0 62, 2010.

\bibitem[Sulyok and Palla(2023)]{sulyok2023greedy}
Bendeg{\'u}z Sulyok and Gergely Palla.
\newblock Greedy routing optimisation in hyperbolic networks.
\newblock \emph{Scientific Reports}, 13\penalty0 (1):\penalty0 23026, 2023.

\bibitem[Krioukov et~al.(2012)Krioukov, Kitsak, Sinkovits, Rideout, Meyer, and
  Bogu{\~n}{\'a}]{krioukov2012network}
Dmitri Krioukov, Maksim Kitsak, Robert~S Sinkovits, David Rideout, David Meyer,
  and Mari{\'a}n Bogu{\~n}{\'a}.
\newblock Network cosmology.
\newblock \emph{Scientific reports}, 2\penalty0 (1):\penalty0 793, 2012.

\bibitem[Clough and Evans(2016)]{clough2016dimension}
James~R Clough and Tim~S Evans.
\newblock What is the dimension of citation space?
\newblock \emph{Physica A: Statistical Mechanics and its Applications},
  448:\penalty0 235--247, 2016.

\bibitem[Bombelli et~al.(1987)Bombelli, Lee, Meyer, and
  Sorkin]{bombelli1987space}
Luca Bombelli, Joohan Lee, David Meyer, and Rafael~D Sorkin.
\newblock Space-time as a causal set.
\newblock \emph{Physical review letters}, 59\penalty0 (5):\penalty0 521, 1987.

\bibitem[Dowker et~al.(2004)Dowker, Henson, and Sorkin]{dowker2004quantum}
Fay Dowker, Joe Henson, and Rafael~D Sorkin.
\newblock Quantum gravity phenomenology, lorentz invariance and discreteness.
\newblock \emph{Modern Physics Letters A}, 19\penalty0 (24):\penalty0
  1829--1840, 2004.

\bibitem[Frohlich(1992)]{frohlich1992non}
J.~Frohlich.
\newblock Regge calculus and discretized gravitational functional integrals.
\newblock In Jurg Frohlich, editor, \emph{Non-Perturbative Quantum Field
  Theory: Mathematical Aspects and Applications}. World Scientific, 1992.
\newblock Selected papers of Jurg Frohlich.

\bibitem[Loll(2019)]{loll2019quantum}
Renate Loll.
\newblock Quantum gravity from causal dynamical triangulations: a review.
\newblock \emph{Classical and Quantum Gravity}, 37\penalty0 (1):\penalty0
  013002, 2019.

\bibitem[Sornette and Ouillon(2012)]{sornette2012dragon}
Didier Sornette and Guy Ouillon.
\newblock Dragon-kings: mechanisms, statistical methods and empirical evidence.
\newblock \emph{The European Physical Journal Special Topics}, 205\penalty0
  (1):\penalty0 1--26, 2012.

\bibitem[Newman(2018)]{newman2018networks}
Mark Newman.
\newblock \emph{Networks}.
\newblock Oxford university press, 2018.

\bibitem[Smechki(2021)]{githubGitHubAntonSemechkoS2SamplingToolbox}
A~Smechki.
\newblock {G}it{H}ub - {A}nton{S}emechko/{S}2-{S}ampling-{T}oolbox: {T}oolbox
  for generating sptially uniform sampling patterns and decompositions of a
  unit sphere --- github.com.
\newblock \url{https://github.com/AntonSemechko/S2-Sampling-Toolbox}, 2021.
\newblock [Accessed 13-07-2024].

\bibitem[Claxton(2006)]{mathworksSurfaceCurvature}
D~Claxton.
\newblock {S}urface {C}urvature --- mathworks.com.
\newblock
  \url{https://www.mathworks.com/matlabcentral/fileexchange/11168-surface-curvature},
  2006.
\newblock [Accessed 13-07-2024].

\end{thebibliography}

\end{document}